\documentclass[journal]{IEEEtran}
\RequirePackage{xcolor}
\usepackage{cite}
\usepackage{amsmath,amssymb,amsfonts}
\usepackage{algorithmic}
\usepackage{graphicx}
\usepackage{textcomp}
\usepackage{wrapfig}
\usepackage{siunitx}
\usepackage{booktabs}
\def\BibTeX{{\rm B\kern-.05em{\sc i\kern-.025em b}\kern-.08em
    T\kern-.1667em\lower.7ex\hbox{E}\kern-.125emX}}
\definecolor{abstractbg}{rgb}{0.89804,0.94510,0.83137}
\begin{document}
\title{Magnetic Localization for In-Body Nano-Communication Medical Systems}
\author{Krzysztof Skos, Albert Diez Comas, Josep Miquel Jornet, \IEEEmembership{Senior Member,~IEEE}, and Pawel Kulakowski
\thanks{This paragraph of the first footnote will contain the date on which you submitted your paper for review. This work was carried out in the framework of COST Action CA20120 INTERACT. It was also supported by the grant CBET-2039189 of US National Science Foundation and by the Polish Ministry of Science and Higher Education with the subvention funds of the Faculty of Computer Science, Electronics and Telecommunications of AGH University. }
\thanks{K. Skos is with the Institute of Telecommunications, AGH University of Krakow, Poland (e-mail: kskos@agh.edu.pl).}
\thanks{A. Diez Comas and J.~M. Jornet are with the Institute for the Wireless Internet of Things, Northeastern University, Boston, MA 02115 USA (e-mails: diezcomas.a@northeastern.edu, j.jornet@northeastern.edu).}
\thanks{P. Kulakowski (corresponding author) is with the Institute of Telecommunications, AGH University of Krakow, Poland (e-mail: kulakowski@agh.edu.pl).}}

\IEEEtitleabstractindextext{%
\fcolorbox{abstractbg}{abstractbg}{%
\begin{minipage}{\textwidth}%
\begin{wrapfigure}[12]{r}{3in}%
\includegraphics[width=3in]{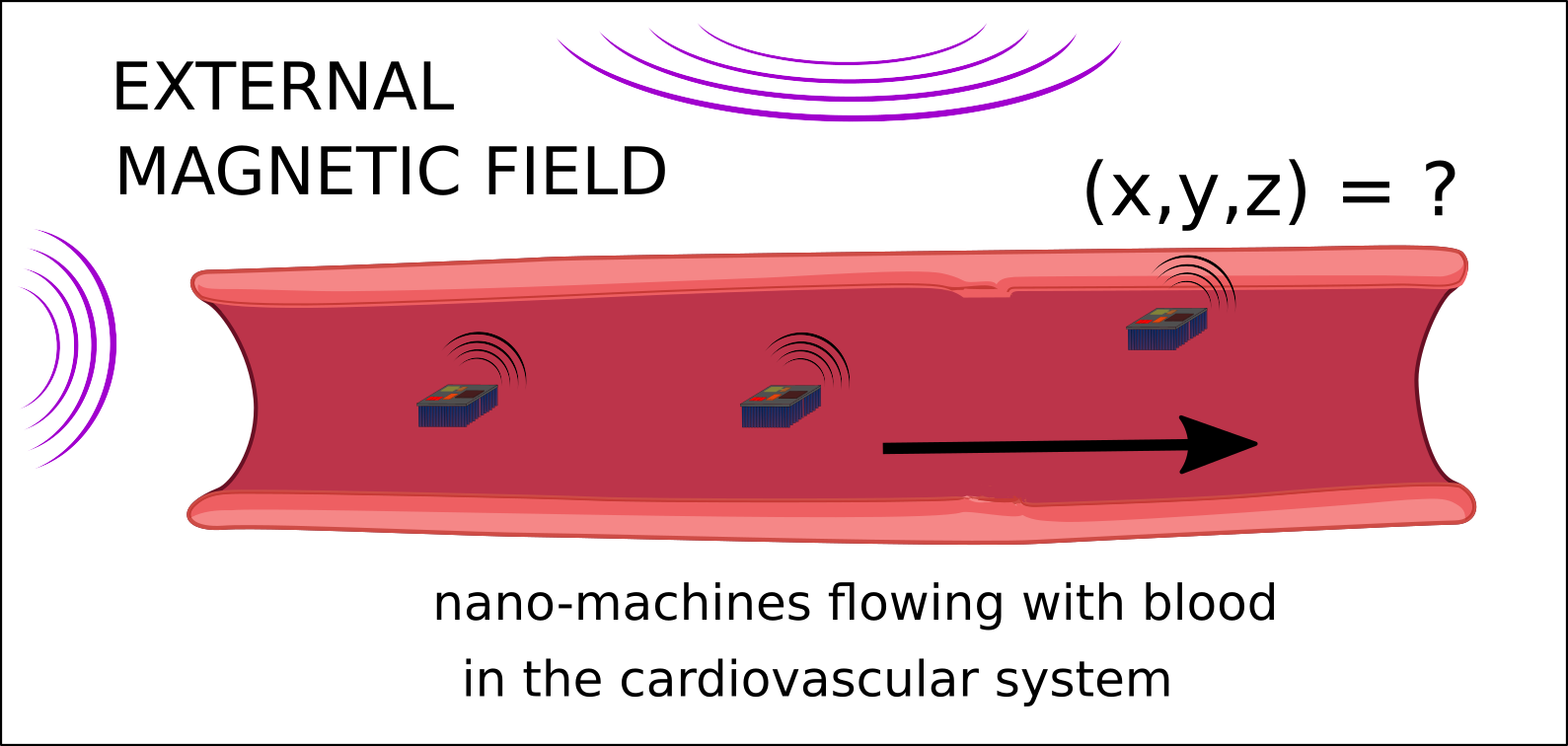}%
\end{wrapfigure}%

\end{minipage}}}

\maketitle

\begin{abstract}
Nano-machines circulating inside the human body, collecting data on tissue conditions, represent a vital part of next-generation medical diagnostic systems. However, for these devices to operate effectively, they need to relay not only their medical measurements but also their positions. This paper introduces a novel localization method for in-body nano-machines based on the magnetic field, leveraging the advantageous magnetic permeability of all human tissues. The entire proposed localization system is described, starting from 10 \si{\micro\meter} $\times$ 10 \si{\micro\meter} magnetometers to be integrated into the nano-machines, to a set of external wires generating the magnetic field. Mathematical equations for the localization algorithm are also provided, assuming the nano-machines do not execute the computations themselves, but transmit their magnetic field measurements together with medical data outside of the body. The whole system is validated with computer simulations that capture the measurement error of the magnetometers, the error induced by the Earth’s magnetic field, and a human body model assuming different possible positions of nano-machines. The results show a very high system accuracy with position errors even below 1 cm. \\ 
\end{abstract}

\section{Introduction}
\label{sec:introduction}
\IEEEPARstart{N}{owadays}, modern medicine is becoming more and more biotechnology-oriented. To have precise information about a human health state, numerous tiny medical devices are placed inside of the body with endoscopy techniques, injection of sensors, or even contrast agents for radiography or magnetic resonance imaging. Some of these diagnostic sensors can even traverse human digestive or cardiovascular systems, gathering valuable information relating to different pathologies. But besides measuring intra-body conditions, to be effective for medical diagnostic purposes, these devices should also have communication and localization capabilities. 

The communication capability is required to deliver the acquired information to medical systems outside of the body. Until now, numerous concepts have been already proposed that integrate tiny medical sensors within a larger scale Internet of Nano-Bio Things~\cite{akyildiz2010internet,Akyildiz2015, Sangwan2022, Aghababaiyan2022, Ali2023}. Assuming the medical sensors are miniature nano-machines, of cubic micrometers in size or even smaller, their communication can be realized according to the so-called nano-communication paradigm, including a few different mechanisms for information transmission~\cite{Kulakowski2020}. The first of them is based on electromagnetic (EM) communication, like in typical wireless systems, but using very high frequencies, usually terahertz and optical bands, suitable for micro-scale devices and short-range information transfer, and built with innovative materials such as graphene and other two-dimensional nano-materials~\cite{Akyildiz2010, Tamagnone2012, Jornet2013, Ukhtary2020}. The second popular approach, called molecular communication, exploits molecules, e.g., waves of calcium ions, vesicles, or bacteria as information carriers~\cite{Akyildiz2008, Farsad2016, Chude-Okonkwo2017}. Other possibilities include very short-distanced but low-delay Forster Resonance Energy Transfer~\cite{Kuscu2015, Solarczyk2016, Kulakowski2017} or the ultrasound (for distances up to a few cm) mechanism~\cite{Hogg2012, Santagati2014}. 

The localization capability, on the other hand, is aimed at finding the exact location of a nano-machine at a chosen moment in time. This is crucial to determine where such a machine takes a measurement, which is particularly important if a specific tissue pathology is detected. The nano-machine might be circulating along blood in the cardiovascular system and its location could constantly change. The localization should be possible not only at the moment of injecting the nano-machine into the body but also later, during its operational time. 

While wireless localization is a well-known and developed topic, the existing solutions do not fit well with the case of in-body nano-machines. First, the tiny nano-machines operate in strictly limited energy budget conditions and are not capable of performing complex computations for their position estimation. Second, the EM propagation conditions are quite harsh, as the EM signals are very strongly attenuated in body tissues. Moreover, the attenuation differs depending on the tissue type, i.e., blood, fat, skin, or bones. This is no better with other nano-communication mechanisms, i.e., molecular, FRET, or acoustic ones. 

Having this in mind, the contribution of this paper is to propose a novel localization approach for in-body nano-machines based on the magnetic field. Like in many localization systems with reference (anchor) points, the reference wires are generating a constant magnetic field. These wires are located in known positions outside of the human body and the emitted magnetic field is constant in time but decreases with the distance from the wires. As the magnetic field penetrates human tissues very well (the relative magnetic permeability is very close to 1 for all human tissues), it can be exploited for determining the distances of a nano-machine from the wires and later the nano-machine position. Moreover, in the proposed solution, the nano-machines do not perform any computations. They just measure the magnetic field, along with other medical measurements. Later, they transmit both the medical and magnetic field measurements wirelessly, e.g., using the THz band~\cite{Jornet2013}, through some intermediate devices to a system outside of the body where all the calculations are performed. We present the whole localization system, starting from the 10 \si{\micro\meter} $\times$ 10 \si{\micro\meter} magnetometers to be integrated into the nano-machines, to the set of wires generating the magnetic field. We provide the equations for the localization algorithm in two versions: for three wires and six or more wires. Then, we assess the localization accuracy with computer simulations taking into account the measurement error of the magnetometers, the error induced by the Earth’s magnetic field, and a human body model assuming different possible positions of nano-machines. The results show a very good system accuracy with localization error even below 1 cm.

The rest of the paper is organized as follows. In Sec.~\ref{sec:related}, we review the related localization approaches.  In Sec.~\ref{sec:magnetic-system}, we present the components of the localization system: nano-machines that can be used for medical diagnostic purposes, micro-magnetometers, and a system of wires for generating magnetic field. Then, in Sec.~\ref{sec:localization-algorithm}, we define the localization algorithm, both its ranging and lateration phases. Later, in Sec.~\ref{sec:simulation-results}, we provide a detailed description of computer simulations validating the localization system and including both the magnetic field generated by the system of wires and the Earth’s magnetic field. We simulate the localization process and we calculate its error for different nano-machine positions according to the accepted 3D human body model. Finally, in Sec.~\ref{sec:conclusions}, we conclude the paper and point out some further research directions.

\section{Related Work on In-body Nano-machine Localization}
\label{sec:related}
Performing wireless localization of mobile nano-machines inside a human body poses several additional challenges compared with the localization in macro-scale wireless networks. First, radio propagation conditions inside the body are quite harsh, with very high signal attenuation, strongly dependent on the type of tissue. For example, for the frequency of 1 THz, which is typically considered for EM nano-nodes, the signal propagation loss at a distance of 1 mm is about 70 dB for fat tissue, 90 dB for skin and even 150 dB for blood~\cite{Canovas2018}, which makes classical receive signal strength (RSS) localization approaches hardly applicable. Second, this extremely high signal attenuation limits the communication range of nano-nodes to about 1-2 mm, so a nano-machine flowing with blood is frequently isolated from other nodes, unless it approaches another nano-machine.

The number of already proposed localization schemes for nano-networks is very limited. Considering nano-machines flowing with blood in a cardiovascular system, an inertial positioning system has been recently proposed~\cite{Simonjan2022}. In this paper, nano-machines are assumed to get information about their estimated position from anchor nodes attached to the skin, when they flow close by. Later, during their movement, nano-machines update their positions based on their velocity and rotation sensors. However, this concept is very challenging, as the blood flow is not laminar all the time, but also turbulent, which makes the position updates inaccurate. Moreover, communication between the skin-attached anchors and nano-machines is rather not possible directly, because of the distance range usually about a centimeter or more. In another paper~\cite{Lemic2020}, the time of flight approach was suggested to calculate the two-way distance between anchors and nano-machines and later to perform localization. This was, though, considered for a grid of static nano-machines. This concept was later developed in~\cite{Lemic2022}, also for static grid-type scenario, where nano-machines further away from the anchors were localized based on the positions of nano-machines closer to the anchors. Two other research contributions~\cite{Tran-Dang2014, Zhou2017} focus on the hop-counting approach, but this is again feasible only for static nano-machines arranged in a grid.

Having in mind these discussed limitations of wireless localization, here, we propose to exploit the properties of a static magnetic field. While an alternating magnetic field, which is associated with electromagnetic waves propagation, is strongly influenced by the type of human tissues it passes through, the situation is different with a static magnetic field. The main advantage of the static magnetic field is its invariability in different types of human tissues. The magnetic permeability, which describes how the static magnetic field penetrates different media, is practically the same, up to the sixth decimal place, in all human tissues~\cite{Collins2002, Greenebaum2018, Duyn2017, Sprinkhuizen2012}, see Tab.~\ref{TablePermeability}. This means that magnetic field strength values can be used for distance estimation independently on the tissues between a magnetic field transmitter and a receiver, e.g., a nano-machine. Moreover, since the relative permeability of various body tissues is close to 1, the static magnetic field experiences minimal disturbance as it moves through a body. It is an additional advantage which facilitates any kind of in-body magnetic field measurements. 

\begin{figure}[tb]
            \centering
            \includegraphics[width=0.95\columnwidth]{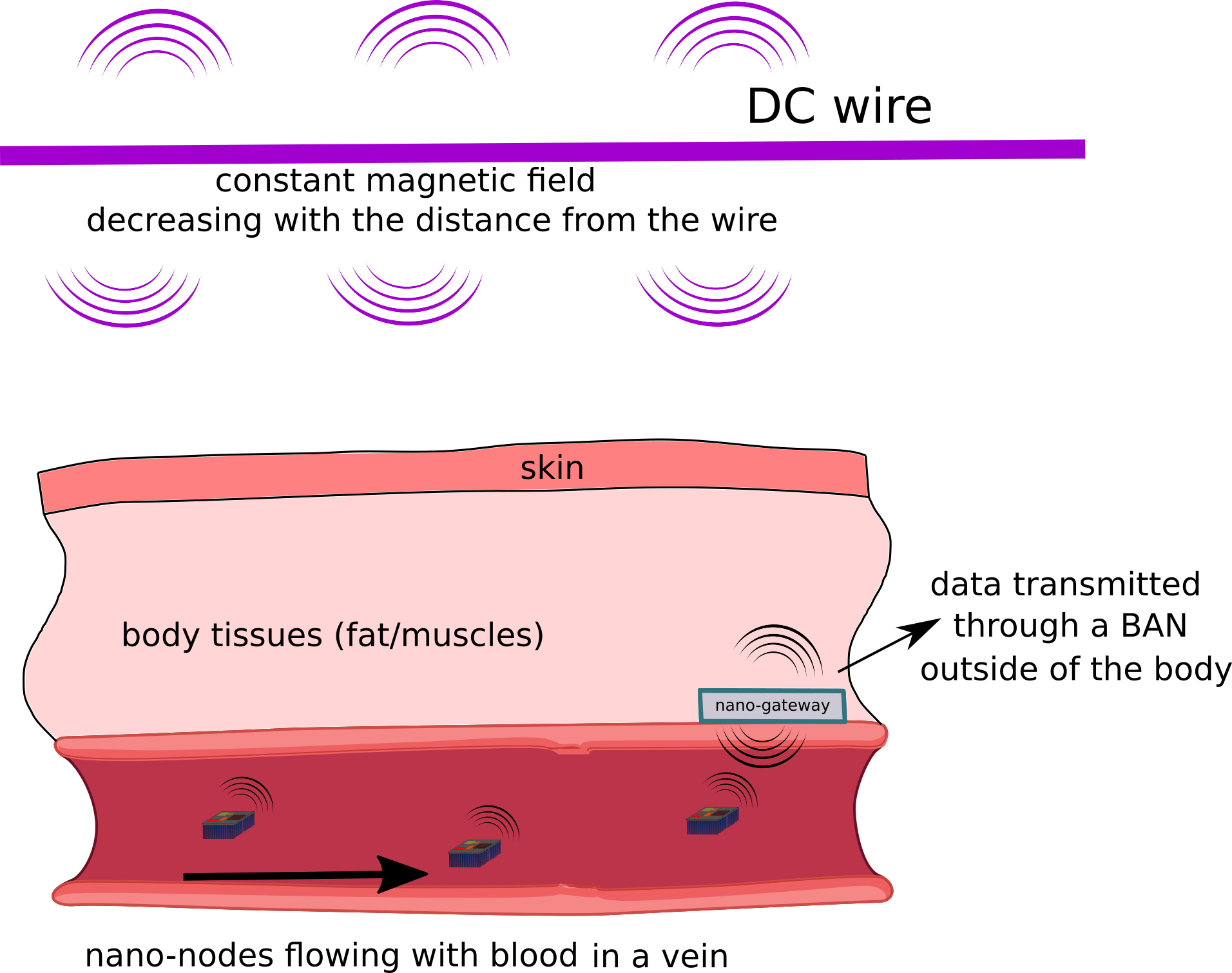}
            \label{fig:magnet-loc-system}
            \caption{The magnetic localization system.}
\end{figure}

To the best of the authors’ knowledge, this is the first paper where magnetic field localization is proposed for nano-machines. There are, however, some published works discussing localization with magnetic fields for networks of larger scale. Localization with the magnetic field generated by external magnets is discussed in~\cite{Emami2018}. The proposed system might be applied inside a body, but the designed mm-size chips transmitting at 480-520 MHz are too large for nano-networks. A system of magnetic coils working as anchors is considered in~\cite{Choi2009, Loke2010, Loke2011}. Also there, the in-body devices are not suitable for nano-networks, as both the magnetic sensor and the wireless transponder are mm-size. Magnetic field localization is proposed as well for applications with  endoscopy capsules being cm-size, assuming a solenoid~\cite{Cavlu2020} or a soft-magnet~\cite{Wen2022} generating the field from inside of a body and some magnetic sensors on the body. In-body magnet rotation can be also tracked for such endoscopy scenarios~\cite{Suveren2019}. Finally, a thorough review of indoor and outdoor magnetic field positioning systems is given in~\cite{Pasku2017}. 

It is worth mentioning that the magnetic field is likewise considered to control the movement of in-body micro-robots, e.g., in~\cite{Sa2023}. Knowing also the popularity of magnetic resonance imaging (MRI) techniques, it seems that soon magnetic field may become a force driving a full spectrum of medical applications for human body inspection and diagnostics.

\section{Magnetic Localization System}
\label{sec:magnetic-system}

Having in mind the main considered application, i.e., in-body nano-networks, here we propose a magnetic field localization system composed of (a) reference anchors outside of a body, and (b) tiny magnetometers mounted on boards of in-body nano-machines. We also assume that information from nano-machines is transmitted outside of the body through some intermediate devices like a nano-gateway. The whole localization process starts with reference anchors emitting the magnetic field. Then, the nano-machines measure the magnetic field with their magnetometers. The nano-machines do not perform any computations. Instead, they wirelessly transmit these readings, together with other medical measurements, through the intermediate devices to the outside of the body. There, the readings are used to calculate the distances of each nano-machine from the wires. Finally, the 3D position of the nano-machine is determined with the aid of the localization algorithm (trilateration or multilateration). This approach keeps the nano-machines simple and avoids wasting their precious energy for localization data processing. The concept is presented in Fig.~\ref{fig:magnet-loc-system} and the main system components are discussed in the three following subsections. 

\begin{table}
\centering
\caption{Relative permeability values \\ for different kind of human tissues ~\cite{Collins2002}}
\begin{tabular}{lcc}
\toprule
         Material &  \textbf{$\mu_r$} \\
\midrule
Free Space    & 1.00000000   \\
Air           & 1.00000040   \\
Water         & 0.99999096   \\
Fat           & 0.99999221   \\
Bone          & 0.99999156   \\
Blood         & 0.99999153   \\
Gray Matter   & 0.99999103   \\
White Matter  & 0.99999120   \\
\bottomrule
\label{TablePermeability}
\end{tabular}
\end{table}

\subsection{Reference Anchors}
\label{sec:reference-anchors}

In localization systems, anchors act as sources of reference signals that enable a localized device to calculate the distances from the anchors. In a magnetic field localization system, the anchors should generate a magnetic field. It has been already suggested to use some coils for this purpose, working like electromagnets~\cite{Loke2010}. However, the mathematical equations of that solution do not have a closed form and they have to be solved numerically with some initial guess values. Instead, here we propose to use a set of metallic wires, each of them supplied with a constant electric current (direct current DC). According to the Biot-Savart law, such a DC wire generates a constant magnetic field decreasing with the distance from the wire axis. Thus, by measuring the magnetic field, a distance from the wire can be determined. It is the basic principle of the localization algorithm proposed here; its mathematical equations are given in Sec.~\ref{sec:localization-algorithm}. We present two versions of the algorithm, the first one for just three wires, and the second one, more robust and accurate, for six or more wires.

\subsection{Nano-machines with Magnetometers}
\label{sec:nano-machines}

Mobile nano-machines form a nano-network operating inside the body, e.g., in the cardiovascular system, collecting medical information relating to different pathologies, from tissue damage to cancer biomarkers~\cite{marzo2019nanonetworks}. Each nano-machine, of size about 10 \si{\micro\meter}$^3$, is a very simple embedded system powered with its own energy source (e.g., a piezoelectric generator) and limited processing, data storage, and communication capabilities~\cite{Akyildiz2010,jornet2023nanonetworking}. 

The miniaturization of an antenna to meet the size requirements of a nano-machine imposes the utilization of very high resonant frequencies. For example, a one-micrometer-long antenna built with a perfect electric conductor (PEC) material would resonate at 150 THz, i.e., in the infrared optical region of the electromagnetic spectrum. There are several challenges associated with the development of optical antennas, starting from the fact that there are no real PEC materials and even very good conductors such as gold or silver have finite complex-valued conductivity, which impacts the antenna design and performance~\cite{alu2007enhanced, nafari2017modeling}. When utilizing optical antennas, on-chip lasers and nano-photodetectors are needed~\cite{feng2014single,nozaki2016photonic}. 

Alternatively, to operate at lower frequencies, graphene-based plasmonic transceivers and antennas could be utilized. Indeed, graphene supports the propagation of surface plasmon polariton waves at terahertz-band frequencies (i.e., between 100 GHz and 10 THz), with very high confinement factors. As a result, a one-micrometer-long graphene-based antenna radiates at approximately 2 THz, i.e., at a frequency nearly two orders of magnitude lower than that of a metallic antenna~\cite{jornet2013graphene,ullah2020review}. When utilizing such antennas, a graphene-based plasmonic nano-transceiver is needed to generate in transmission and process in reception THz signals~\cite{tredicucci2013device,crabb2021hydrodynamic}. If even lower frequencies are needed, recently, magneto-electric antennas able to operate at hundreds of MHz have been proposed~\cite{nan2017acoustically,zaeimbashi2019nanoneurorfid}. In all these cases, individual elements have been developed, and their integration in a miniature nano-radio is the next step. In~\cite{abadal2024electromagnetic}, a good summary of the state of the art is provided.

When it comes to the energy management system of nano-machines, the situation is quite similar. Due to the nature of nano-machines and the inability to manually replace or recharge their batteries, energy harvesting systems are critical~\cite{jornet2012joint}. Different technologies have been proposed to collect energy from the environment. The most commonly cited relies on piezoelectric nanogenerators built with zinc oxide nanowires that harvest energy from the heartbeat or the blood flow~\cite{hu2019strategies}. There have been several studies aimed at optimizing the performance of self-powered nano-machines~\cite{yao2017achievable}.

In this paper, we assume the nano-machines, in addition to medical sensors, are equipped with magnetometers, i.e., instruments to measure the magnetic field. In general, magnetometers are large-scale devices that can be classified into many types like magnetoresistive, spin-valve, superconducting quantum interference devices, or Hall effect-based. For nano-machines, the magnetometer size should be much smaller than most of them, so here we propose to take advantage of a very recently proposed solution: a graphene Hall effect magnetometer~\cite{Izci2018}. The active area of this magnetometer is 10 \si{\micro\meter} x 10 \si{\micro\meter}, which is comparable with state-of-the-art nano-machines and hopefully could be even reduced with the progress in this technology. This type of device was already tested experimentally, reporting a steady relative measurement error of 1\% for the magnetic field from 0 to 120 mT, with the step of 2 mT~\cite{Izci2018}.

The magnetic field is a vector parameter and a single Hall effect magnetometer measures just a single component: $x$, $y$, or $z$. Thus, each nano-machine is equipped with three such devices, together measuring all three field components. 

In addition to the magnetometers, a control unit, a communication unit and an energy management system, at the very least, are needed. Aligned with the vision of nano-machines~\cite{Akyildiz2010,jornet2023nanonetworking,abadal2024electromagnetic}, the size of the envisioned embedded nano-systems should not exceed tens to a few hundred cubic micrometers. The energy management nano-system, often considered to be based on piezoelectric or acoustic harvesting technologies, has usually been the dominating element. Instead, compact on-chip radios based either on plasmonic or magnetoelectric antennas are significantly smaller (sub-micrometric). With current experimentally developed solutions, the magnetometers will be comparable in size to the energy harvesting nano-system. This motivates further research on nano-systems integration.

When performing a medical measurement that requires additional information where the measurement is taken, the magnetometers’ readings are saved together with the medical ones. Later, all this data is transmitted out of the body through the intermediate devices. The nano-machine position is calculated after that, in a medical data center or a doctor's computer, so that the nano-machine processor is not burdened with these calculations. 

\subsection{Intermediate Devices}
\label{sec:intermediate}

As mentioned in Sec.~\ref{sec:related}, the tiny nano-machines circulating in a human body have very restricted capabilities, with a communication range limited to about 1-2 mm. Because of that, a nano-machine flowing through the same vein where a nano-gateway is installed has only a limited probability of successful transmission to this device~\cite{Asorey2020}. This topic was already further investigated for so-called flow-guided nano-networks in numerous medical applications like diagnosis of artery occlusion~\cite{Asorey2023}, bacterial infections, sepsis, heart attacks, or restenosis~\cite{Canovas2020}. The results show that with a sufficient number of nano-machines in the network, the probability of a successful information transfer out of the body is close to 100\%, but it may take even a few hours~\cite{Asorey2023}. A recent work~\cite{Garcia2023} shows that multi-hop communication between the nano-machines may additionally improve this information transfer. While the detailed analysis of this communication is out of the scope of this paper, we assume that nano-machines can deliver their gathered medical and magnetic field data to a medical data center through intermediate devices like a nano-router and a body area network. This communication process is assumed to be performed after the localization procedure, which takes no longer than 1 ms (see Sec.~\ref{sec:results}).

\section{Localization Algorithm}
\label{sec:localization-algorithm}

In this section, we explain the proposed localization algorithm. Our proposed solution shares some similarities with the receive signal strength approach, but here the magnetic flux density is measured instead of the radio signal strength. The proposed algorithm has two phases: ranging and lateration. While the ranging is the same for different arrangements of DC wires, the lateration phase depends on the number of wires. The basic approach requires only three wires, one for each plane. The more complex algorithm requires at least six wires, three in one plane and three in another plane. 

\subsection{Ranging}
\label{sec:ranging}
In the ranging phase, each DC wire is activated in a sequence. At any moment, only one wire is activated. First, nano-machines measure the magnetic flux density generated by all DC wires, one wire at a time. Since each nano-machine has three magnetometers, three measurements are taken simultaneously for each wire, corresponding to three orthogonal components of the flux density. These three measurements are then combined into a single vector, whose magnitude is calculated. Finally, the distance from the wire is calculated with the Biot-Savart law:
\begin{equation}
    R = \frac{\mu I}{2\pi B},
    \label{eq:Biot-Savart}        
\end{equation}
where
\begin{itemize}
    \item $R$ - distance between the nano-node and the wire
    \item $I$ - electric current flowing through the wire
    \item $B$ - magnitude of measured magnetic flux density
    \item $\mu$ - magnetic permeability of the medium.
\end{itemize}

\subsection{Trilateration with Minimal Number of Wires} 
\label{sec:alg3}
\begin{figure}
            \centering
            \includegraphics[width=0.32\textwidth]{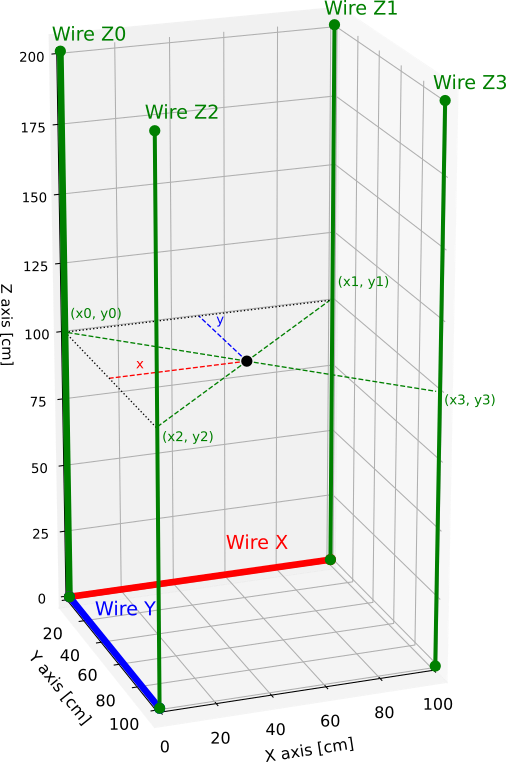}
            \caption{For localization, the minimal set of wires consists of X, Y and Z0 wires (see Sec. ~\ref{sec:alg3}). A more robust approach (see Sec. ~\ref{sec:alg15}) requires multiple wires parallel to each of the axes; here four Z wires are shown: Z0-Z3.}
            \label{fig:wires}
        \end{figure}

In this scenario, the system is considered to have the minimal required number, i.e., three wires. The wires are aligned along the X, Y, and Z axes of the coordinated system (see the wires X, Y, and Z0 in Fig.~\ref{fig:wires}). The nano-machines are located in the (+x, +y, +z) octant of the coordinate system, so all three coordinates are always positive values. Each distance between a nano-machine and a wire, calculated in the ranging phase, creates a cylinder of possible nano-machine positions along the wire. For each wire, only two coordinates are involved. Thus, with three wires, a set of simple three equations can be formed, one for each plane:  
\begin{equation}\label{eq:Rx2}
    \begin{split}
        R^2_x &= y^2 + z^2 \\
        R^2_y &= x^2 + z^2 \\
        R^2_z &= x^2 + y^2,
    \end{split}
\end{equation}
where
\begin{itemize}
    \item $x$, $y$, $z$ - unknown coordinates of the nano-machine
    \item $R_x$, $R_y$, $R_z$ - distances from the nano-machine to the wires X, Y, Z, respectively.
\end{itemize}

As the coordinates are positive values, they can be calculated from \eqref{eq:Rx2} as: 
\begin{equation}\label{eq:Rx3}
    \begin{split}
        x &= \sqrt{\frac{1}{2}(R^2_z + R^2_y - R^2_x)} \\
        y &= \sqrt{\frac{1}{2}(R^2_x + R^2_z - R^2_y)} \\
        z &= \sqrt{\frac{1}{2}(R^2_x + R^2_y - R^2_z)}.
    \end{split}
\end{equation}

While this approach theoretically can provide the nano-machine position, in practice, it is not robust against measurement errors. When a nano-machine is located close to one of the $XY$, $XZ$, or $YZ$ planes (see Fig.~\ref{fig:wires}), one of the estimated distances might be much larger than the other ones. Consequently, one of the root values in~\eqref{eq:Rx3} might be negative. It is then set to zero, to avoid false imaginary parts of the calculated coordinates. 

\subsection{Multilateration with Many Wires} 
\label{sec:alg15}
A more robust algorithm requires a larger number of reference wires. In the following section, we present an approach feasible when we have at least six wires. Let us consider that all the wires are parallel to one of the axes of the coordinates system and call them $X$, $Y$, or $Z$ wires, according to the axis of their parallelism. 
        
Now, let us consider a set of parallel wires, e.g., $Z$ wires, like in Fig.~\ref{fig:wires}. Again, the distances calculated in the ranging phase create cylinders of possible nano-machine positions along each wire. Thus, localization with Z wires enables to calculation of the ($x_z$,$y_z$) coordinates of the nano-machine with the following set of equations:
\begin{equation}\label{eq:Lange1}
            \begin{cases}
                (x_z - x_1)^2 + (y_z - y_1)^2 = R^2_{z1}\\
                (x_z - x_2)^2 + (y_z - y_2)^2 = R^2_{z2}\\
                ... \\
                (x_z - x_n)^2 + (y_z - y_n)^2 = R^2_{zn}
            \end{cases}
        \end{equation}
where
\begin{itemize}
            \item $n$ - number of Z wires
            \item $x_z$, $y_z$ - coordinates of the nano-machine estimated with Z wires
            \item $R_{zi}$ - distance from the nano-machine to the wire $Z_i$
            \item ($x_i$, $y_i$) - wire $Z_i$ coordinates.
        \end{itemize}

Considering the first $Z$ wire is positioned exactly along the $z$ axis, its coordinates are $x_0$=0 and $y_0$=0. Then, following the approach presented in~\cite{Langendoen2003}, and assuming we have at least three $Z$ wires, the first equation in~\eqref{eq:Lange1} can be subtracted from the rest of the equations, and after reordering the terms, it results in:
\begin{equation}\label{eq:matri}
            \begin{cases}
                x_2 x_z + y_2 y_z = \frac{1}{2} (R^2_{z1} - R^2_{z2} + x^2_2 + y^2_2) \\
                \qquad \qquad \qquad ...\\
                x_n x_z + y_n y_z = \frac{1}{2} (R^2_{z1} - R^2_{zn} + x^2_n + y^2_n).
            \end{cases}
        \end{equation}
This set of equations can be presented in a matrix form:
\begin{equation}
            \begin{split}
                \begin{bmatrix}
                    x_2 & y_2 \\
                    ... & ... \\
                    x_n  & y_n 
                \end{bmatrix}
                \begin{bmatrix}
                    x_z \\
                    y_z
                \end{bmatrix}
                = \frac{1}{2}
                \begin{bmatrix}
                    R^2_{z1} - R^2_{z2} + x^2_2 + y^2_2 \\
                    ... \\
                    R^2_{z1} - R^2_{zn} + x^2_n + y^2_n.
                \end{bmatrix}
            \end{split}
        \end{equation}
Denoting the matrices as
\begin{equation}
            A = 
            \begin{bmatrix}
                    x_2 & y_2 \\
                    ... & ... \\
                    x_n  & y_n 
            \end{bmatrix}
            , \qquad r = 
            \begin{bmatrix}
                    x_z \\
                    y_z
            \end{bmatrix}
        \end{equation}

        \begin{equation}
            b = \frac{1}{2}
            \begin{bmatrix}
                    R^2_{z1} - R^2_{z2} + x^2_2 + y^2_2 \\
                    ... \\
                    R^2_{z1} - R^2_{zn} + x^2_n + y^2_n,
            \end{bmatrix}
        \end{equation}
Equation \eqref{eq:matri} can be simply written as:
\begin{equation}
            A r = b.
\end{equation}
Then, this matrix equation can be solved using the standard least-squares approach, obtaining the estimate vector $r$ containing the $x_z$ and $y_z$ coordinates of the nano-machine:
\begin{equation}\label{eq:final}
            r = (A^T A)^{-1} A^T b.
\end{equation}

Now, the analogous approach can be performed for $X$ and $Y$ wires. Having at least three $X$ wires and three $Y$ wires, we can follow the algorithm given in~\eqref{eq:Lange1}-\eqref{eq:final} to obtain ($y_x$,$z_x$) and ($x_y$,$z_y$) coordinates. Each set of wires enables to calculation of two of the nano-machine coordinates only. However, assuming the number of $X$, $Y$, and $Z$ wires is respectively equal to $n$, $m$, and $p$, the final nano-machine coordinates can be calculated as respective weighted means:
\begin{equation}\label{eq:coordinates}
            \begin{split}
                x &= \frac{mx_y + px_z}{m+p}\\
                y &= \frac{ny_x + py_z}{n+p}\\
                z &= \frac{nz_x + mz_y}{n+m}.
            \end{split}
        \end{equation}

Here, it should be noted that it is sufficient to have just 6 wires in total, e.g, three $X$ wires, three $Z$ wires and no $Y$ wires. It means that $m$ is equal to 0 and the parameters $x_y$ and $z_y$ cannot be calculated, but according to~\eqref{eq:coordinates} all three coordinates of the nano-machine can be estimated. Thus, the minimum number of wires for this version of the localization algorithm is equal to 6.

\section{Simulation Results}
\label{sec:simulation-results}
\begin{figure}
    \centering
            \includegraphics[width=0.35\textwidth]{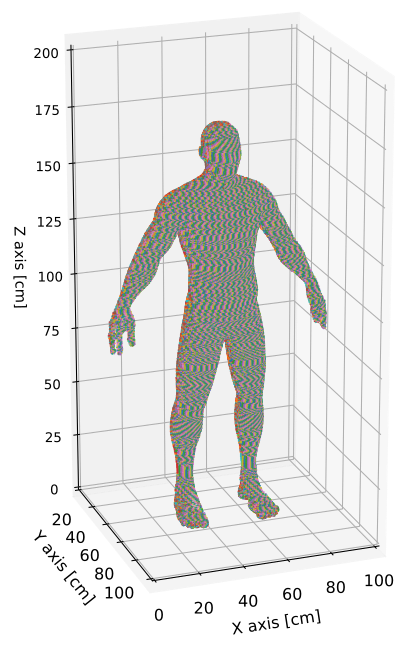}
            \caption{The human body model.}
            \label{fig:body_model}
\end{figure}

To validate the proposed localization algorithms, a computer simulator is written in Python 3 programming language. The simulator reflects the geometric setting of DC wires being the magnetic sources (Sec.~\ref{sec:reference-anchors}) and nano-machines equipped with magnetometers operating inside of a human body (Sec.~\ref{sec:nano-machines}). It also takes into account the Earth's magnetic field which might be a source of measurement errors (Sec.~\ref{sec:earth-field}). For each position of the nano-machine inside of the body, the simulator calculates the measured magnetic field coming from each DC wire influenced by the Earth's magnetic field. With these measurements, the position of the nano-machine is estimated based on the proposed localization algorithm (Sec.~\ref{sec:localization-algorithm}). Finally, the localization error is calculated in comparison to the real position of the nano-machine. Discussing it in more detail, each simulation follows the subsequent steps:
\begin{enumerate}
    \item \textbf{Geometry definition.} The general geometry of the simulation is created: the position of the DC wires and the relative inclination of the Earth's magnetic field are established. It should be noted that the Biot-Savart law~\eqref{eq:Biot-Savart} is defined for infinitely long DC wires. In order to have accurate ranging estimations, the wires should be much longer than the human body, depending on the required accuracy even a few tens of meters long.  
             
    \item \textbf{Nano-machines positions and orientations.} A voxel model of a 175~cm human male is implemented to emulate the human body. The model is depicted in Fig.~\ref{fig:body_model}; it is composed of 625,885 points in total, which results in the point spatial resolution of 0.5 cm along each axis. Each point represents a potential nano-machine position for which magnetic field measurements are performed and localization is attempted. At each point, a nano-machine is generated with a random orientation, which is repeated 100 times for statistical credibility.	
            
    \item \textbf{Magnetic field of DC wires.} For the chosen electric current, the magnetic flux density generated by a DC wire is calculated at each point of the human body, according to the Biot-Savart law~\eqref{eq:Biot-Savart}. Then, the measurement noise is added, which is generated from a uniform distribution of 1\% of the computed flux density. The whole step is repeated for each DC wire generating the magnetic field. 
            
    \item \textbf{Earth's magnetic field error.} A simulated error, resulting from the difference between the real Earth's magnetic field and its modeled value, is added to each measured magnetic flux density. The Earth's field model with its accuracy are discussed in a separate Section \ref{sec:earth-field}. 
            
    \item \textbf{Localization and its error.} The proposed localization algorithm is executed and the position of the nano-machine is estimated. The Euclidean distance between the real and estimated positions is calculated as the position error. It is averaged over 100 random nano-machine orientations at each point of the human body model. 
\end{enumerate}

\subsection{Earth's Magnetic Field and Other Sources of Interference}
\label{sec:earth-field}
In static magnetic field measurements, the Earth's magnetic field (also called geomagnetic field) must be considered. This field is created naturally in the so-called geodynamo process. As a result, Earth is a large magnetic dipole, and its poles are slowly changing their positions in time. The average magnetic flux density at the Earth's surface is between 25 and 65~\si{\micro\tesla}. 
        
Ignoring the fact that Earth’s magnetic field exists and performing the localization would result in huge position estimation errors. We performed computer simulations to check this approach and the obtained position errors were about several dozen centimeters, which is at least an order of magnitude larger than the errors reported in Sec.~\ref{sec:results}. Fortunately, a couple of models allow for the estimation of the geomagnetic field. In this paper, we use the commonly known World Mathematical Model (WMM)~\cite{wmm2020}. It was created by the U.S. National Oceanic and Atmospheric Administration’s National Centers for Environmental Information (NOAA/NCEI) and the British Geological Survey (BGS). The current model version was parameterized in 2020 and is supposed to be valid until 2025. The main goal of the WMM is to represent the planet's magnetic field for all locations of the globe. What is quite important here, it also provides data on how large the model error is compared with the real Earth's magnetic field. These errors for three orthogonal components (northern, eastern, and vertical ones) of the magnetic flux density are shown in Tab.~\ref{Table2} for the 2020 and 2025 years. 

For the computer simulator, we calculate the average of the model error for 2020 and 2025 (see the third column of Tab. \ref{Table2}). Then, for each nano-machine, we randomly generate the WMM errors for all three components as uniformly distributed values from the range $\langle$-error, +error$\rangle$. Thus, the final error of the measured magnetic field results from: (1) the WMM error, and (2) the relative measurement error of the magnetometers being 1\% (see Sec.~\ref{sec:nano-machines}).

Except for Earth's magnetic field, some other sources can create interference for the magnetic field measured by magnetometers. The vast majority of magnetic field sources generate an alternating magnetic field; these are sources of electromagnetic waves or just devices powered with electrical AC current. It is assumed here that magnetometers are equipped with filters that pass a static magnetic field only. There are, however, some sources of static magnetic field, mostly cables with DC current. A low DC current is not going to create a significant magnetic field, as its field density decreases with distance. Nonetheless, some medical equipment like MRI devices can be supplied with a DC current even above 80 A, and thus placing the discussed localization system close to MRI devices, e.g., in the same room, should be clearly avoided. 

\begin{table}
\label{tab:wmm_error}
        \caption{Estimated WMM error and average variances}
        \begin{tabular}{lccc}
        \toprule
        Earth's field &  WMM error   &  WMM error   &  Computer simulator \\ 
        component & in 2020 & in 2025 & error \\
        \midrule
        X (northern) [nT]  &  127  &  135  &  131 \\
        Y (eastern) [nT]  &  86  &  101  &  94 \\
        Z (vertical) [nT]  &  146  &  168  &  157\\
        \bottomrule
        \label{Table2}
        \end{tabular}
\end{table} 

\begin{figure}[tb]
    \centering
    \includegraphics[width=0.32\textwidth]{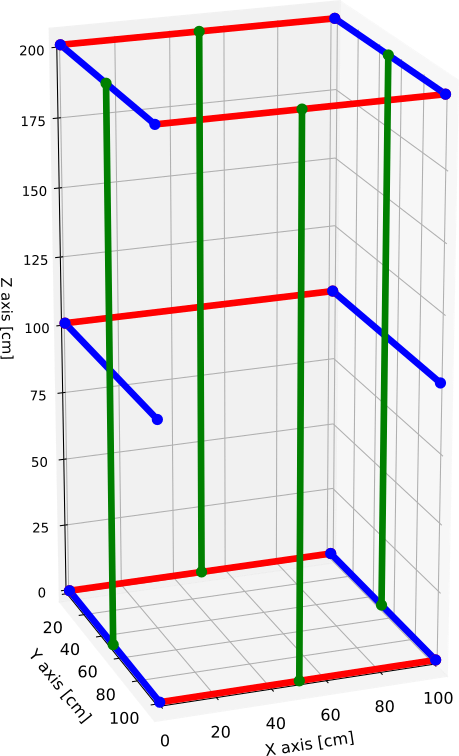}
    \caption{The 15-wires arrangement: wires $X$ (red), wires $Y$ (blue), and wires $Z$ (green).}
    \label{fig:magnets15}
\end{figure}
    
\subsection{Results}
\label{sec:results}
Several wire arrangements were investigated during simulations. The first one was the case with just 3 DC wires, as it is the minimal number of wires when the localization is feasible with the trilateration approach (Sec.~\ref{sec:alg3}). Then, scenarios with 6, 9, 15, and 30 DC wires are considered, for the more advanced algorithm, based on multilateration (Sec.~\ref{sec:alg15}). These wire arrangements create cage-like shapes where a patient can stand inside (see an example for 15 wires in Fig.~\ref{fig:magnets15}) for medical investigation. The electrical current for all the wires was 100 A. The created magnetic flux density was never higher than 120 mT inside the human body, as it was the maximum value in the measurement range of the used magnetometers. Also, one scenario with a lower electrical current, only 10 A, was investigated; it resulted in a proportionally lower magnetic flux density inside the body. For each scenario and each of 625,885 nano-machine positions, 100 simulations were carried out and the average localization error was calculated. After performing Kolmogorov-Smirnov tests, it was noticed that, over different nano-machine positions, the localization errors did not have Gaussian distributions. Thus, their values are given with 1st, 2nd (median), and 3rd quartiles, together with maximum values. All these statistics are shown in Tab.~\ref{tab:loc_error}.

It should be considered that nano-machines might move during the localization process. As it is assumed the nano-machines flow with the blood, their typical velocity is about 25 cm/s or lower, but in some rare cases can be even 50 cm/s. The time duration of the localization process mainly depends on the time of switching on and off the wires, as each wire should generate the static magnetic field separately. This issue was investigated performing additional calculations with COMSOL Multiphysics software. These investigations showed that even for the current of 100 A, the current stabilized after the time of about 1 ~\si{\micro\second}. Thus, it might be assumed that the whole process of switching on, performing the measurements and switching off for even 30 wires is not going to be longer than 1 ms. Even for the rare occurring velocity of 50 cm/s, a nano-machine does not move further than 0.5 mm in that time period. This movement error is then treated as not significant, compared with localization errors reported below, being at least an order of magnitude larger.

\begin{figure}[tb]
                \centering
                \includegraphics[width=0.45\textwidth]{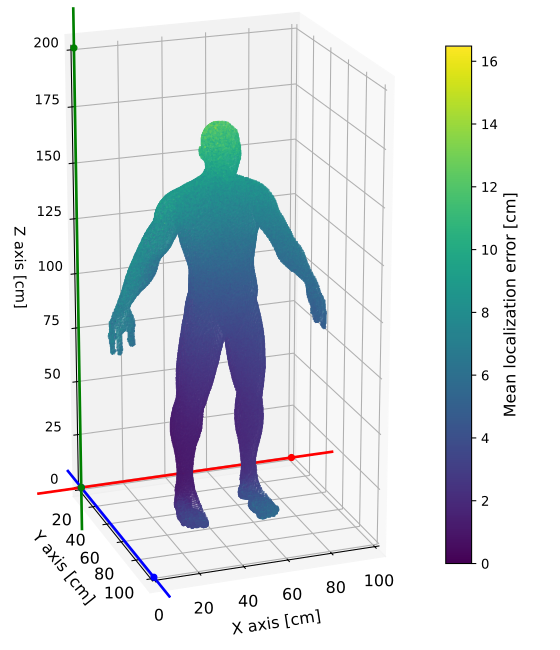}
                \caption{Map of mean localization error in cm for 3 wires scenario.}
                \label{fig:results3}
        \end{figure}

In the 3 wires scenario, the median position error equals 5.01 cm, with the 1st and 3rd quartiles being 2.97 and 7.16 cm, respectively. The detailed map of the median error is depicted in Fig.~\ref{fig:results3}. The error increases in the regions being far from the wires, especially the head, which is located about 170 cm from both $X$ and $Y$ wires. It is due to the lower magnetic field there when generated by these wires, thus the disturbances resulting from the Earth’s field (see Sec.~\ref{sec:earth-field}) are more significant. 

\begin{figure}[tb]
                \centering
                \includegraphics[width=0.45\textwidth]{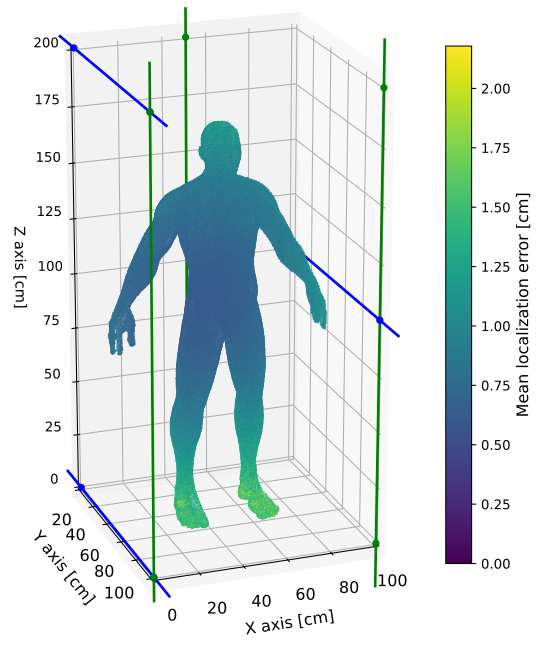}
                \caption{Map of mean localization error in cm for 6 wires scenario.}
                \label{fig:results6}
        \end{figure}

For the multilateration algorithm, the scenarios with 6, 9, 15, and 30 wires are compared in Tab.~\ref{tab:loc_error}. The reported errors are much lower than for only 3 wires. Especially the scenario of 6 wires, which is the minimum number for multilateration, seems promising: the median position error is just 0.79 cm, with the 1st and 3rd quartiles being 0.7 and 0.95 cm, respectively. This proves the advantage of the multilateration approach over the simplistic trilateration. The error map for the 6 wires scenario is presented in Fig.~\ref{fig:results6}. Here, the error is the highest in the feet area. It is because of the wires’ location: they are positioned around the body, but the feet are a body part farthest from the center of the wires’ system. In practical applications, it might be possible to adapt the size of the wires’ system to the height of the body decreasing the localization error. It might be also worthwhile to optimize the exact location of the wires. They should be positioned close to the body, to have a strong magnetic field there. Because of this reason, in this scenario, it was important to use a set of three $Z$ wires, which are located along the body, and one of the other set of perpendicular ones, e.g. $Y$ wires, as required for the localization algorithm. Probably, it is possible to obtain a slightly better localization accuracy after a careful optimization of the wires’ positions. This might be done e.g. with gradient methods, but it depends on the exact shape of the body and it is out of the scope of this study. 

Further increasing the number of wires improves the localization accuracy only slightly. For 9 wires, the median error (2nd quartile) is 0.77 and it remains the same, accurate to two decimal places, for 15 wires. In the scenario with 30 wires, the median error decreases to 0.69 cm, so there is an improvement, but it is just a 13\% lower error compared with 6 wires. 

Finally, a case of a lower DC current was investigated. Two scenarios were compared for 15 wires: apart from the current of 100 A, a case of 10 A was simulated as well (see Tab.~\ref{tab:loc_error}). For the lower current in the wires, the localization error significantly increases. The median position error grows from 0.77 cm (100 A) to 3.05 cm (10 A). This situation can be explained by the disturbances created by the Earth’s magnetic field. The lower DC current in the wires results in a lower magnetic field density, thus the error of the WMM (Sec.~\ref{sec:earth-field}) model is relatively stronger. 

        \begin{table*}
        \centering
        \caption{Position error summary}
        \begin{tabular}{lcccccc}
        \toprule
        Scenario  &  3 wires, 100 A & 6 wires, 100 A & 9 wires, 100 A & 15 wires, 100 A & 15 wires, 10 A & 30 wires, 100 A\\
        \midrule
        Errors [cm]: & & & & & &   \\
        Maximum X error         &  10.88 &  0.97  & 1.00 & 0.89 & 4.80 & 0.76 \\
        Maximum Y error         &  7.63  &  0.50  & 0.92 & 0.94 & 2.40 & 0.93 \\
        Maximum Z error         &  9.78  &  1.88  & 1.55 & 1.29 & 10.64 & 1.25 \\
        Maximum position error  &  16.48 &  2.18  & 2.15 & 1.78 & 11.11 & 1.75 \\
        Median X error          &  2.12  &  0.38  & 0.38 & 0.37 & 0.83 & 0.31 \\
        Median Y error          &  1.71  &  0.24  & 0.38 & 0.39 & 0.87 & 0.36 \\
        Median Z error          &  3.48  &  0.56  & 0.42 & 0.36 & 2.46 & 0.33 \\
        Median position error   &  \textbf{5.01}  &  \textbf{0.79}   & \textbf{0.77} & \textbf{0.77} & \textbf{3.05} & \textbf{0.69} \\
        1\textsuperscript{st} quartile X error  &  1.40   &  0.33   & 0.34 & 0.30  & 0.53 & 0.24 \\
        1\textsuperscript{st} quartile Y error  &  1.04  &  0.22   & 0.33 & 0.33 & 0.61 & 0.29 \\
        1\textsuperscript{st} quartile Z error  &  2.07  &  0.46   & 0.34 & 0.26 & 1.18 & 0.22 \\
        1\textsuperscript{st} quartile position error  &  2.97  &  0.70   & 0.68 & 0.60 & 1.82 & 0.52 \\
        3\textsuperscript{rd} quartile X error  &  3.23   &  0.45   & 0.45 & 0.46 & 1.28 & 0.40 \\
        3\textsuperscript{rd} quartile Y error  &  2.82   &  0.26   & 0.44 & 0.48 & 1.12 & 0.47 \\
        3\textsuperscript{rd} quartile Z error  &  4.91   &  0.69   & 0.55 & 0.49 & 3.84 & 0.47 \\
        3\textsuperscript{rd} quartile position error  &  7.16  &  0.95   & 0.92 & 0.94 & 4.33 & 0.87 \\
        \bottomrule
        \label{tab:loc_error}
        \end{tabular}
        \end{table*}

\section{Conclusions}
\label{sec:conclusions}

In this paper, we presented a new magnetic-field-based localization system suitable for in-body nano-machines. In the system, external electric wires with constant current generate a magnetic field that can be measured by tiny magnetometers mounted on board the nano-machines. These measurements are sent out of the body through a body-area network and are used to calculate the nano-machine positions. We proposed two versions of the localization algorithm: for three, and six or more electric wires. We also explained how to take into account Earth's magnetic field. We performed computer simulations for both algorithm versions showing that even with errors induced by the magnetometers and Earth's magnetic field, the obtained localization accuracy can be about 1 cm or better. The proposed magnetic localization approach is a solution that fits well with the currently increasing number of in-body medical systems. Recently, micro- and nano-scale robots for drug delivery are gaining attention, together with techniques where the magnetic field is used for their control and locomotion~\cite{Murali2022}. It can also observe the development of systems based on ingestible electronics proposing an opportunity to monitor gastrointestinal diseases in real time~\cite{Abdigazy2024}. For all these devices, the presented localization system can offer their precise position estimation, expanding their capabilities and applications in personalized medical treatment. 

Future works extending these studies may go in many different directions ranging from specific system applications through nano-device hardware fine-tuning to new networking architectures. One particularly interesting example is substituting the external DC wires, generating the magnetic field, with in-body magnets. Such magnets are already considered for guiding some nano-particles~\cite{Blumler2021}. The advantage of this solution is the fact that no external reference wires are required, as the source of the magnetic field is inside the body. It is, however, still to be investigated if the magnets' positions could be sufficiently stable and the fields suitably regular and strong for accurate localization. Another related research line is a study on generating a uniform magnetic field around external wires of finite length. Here we must assume the wires are quite long, even a few tens of meters, but for practical reasons it would be more convenient to use shorter wires. Thus, it would be worth examining what is the localization error resulting from the non-uniform magnetic field and how a uniform field can be generated in a more compact wires’ set-up. When considering a specific number of wires and a particular human body, the wires’ exact positions could be optimized, e.g. with gradient methods, for the most uniform magnetic field and the best localization performance (moving the wires closer to the body center and adapting their positions to the body shape). These examples of future studies just indicate a wide range of research topics related to magnetic field applications for in-body medical diagnosis.

\section*{Acknowledgement}
The authors would like to thank Andrzej Kozłowski for his invaluable comments and suggestions regarding magnetometers.

\bibliographystyle{IEEEtran}
\bibliography{references}

\end{document}